\documentclass[twocolumn,twocolappendix,appendixfloats]{aastex6}
\usepackage{subfigure}
\usepackage{caption}
\usepackage{threeparttable}
\usepackage{multirow}
\usepackage{remreset}
\usepackage{CJKutf8}
\usepackage{rotating} 
\collaborationName{Friends of AASTeX}
\shorttitle{Correlations between the variation of the continuum and BALs}
\shortauthors{Lu \& Lin}
\begin{document}


\title{Correlations between the variation of the ionizing continuum and broad absorption lines in individual quasars}

%% Use \author, \affil, and the \and command to format author and affiliation 
%% information.  If done correctly the peer review system will be able to
%% automatically put the author and affiliation information from the manuscript
%% and save the corresponding author the trouble of entering it by hand.
%%
%% The \affil should be used to document primary affiliations and the
%% \altaffil should be used for secondary affiliations, titles, or email.

%% Authors with the same affiliation can be grouped in a single
\author{
\begin{CJK*}{UTF8}{gbsn}
Wei-Jian Lu (陆伟坚)\altaffilmark{1} and Ying-Ru Lin (林樱如)\altaffilmark{2}
\end{CJK*}
}
\affil{School of Information Engineering, Baise University, Baise 533000, China}

\altaffiltext{1}{E-mail: william\_lo@qq.com (W-J L)}
\altaffiltext{2}{E-mail: yingru\_lin@qq.com (Y-R L)}

%% AASTeX 6.0 supports the ability to suppress the names and affiliations
%% of some authors and displaying them under a "collaboration" banner to
%% minimize the amount of author information that to be printed.  This 
%% should be reserved for articles with an extreme number of authors.  
%% The necessary command are \AuthorCallLimit and \collaborationName.
%% An \AuthorCallLimit=2 call prior to the author list will only show
%% the authors in the first two \author calls.  The \collaborationName
%% defines the collaboration identifier.  Commented examples below.

%\AuthorCallLimit=1
%% Will only show Schwarz & Muench since Schwarz and Muench
%% are in the same \author call. 
%\collaborationName{Friends of AASTeX}
%% will print "The AAS collaboration" after the shortened author list.
%% Note that all the \altaffil information will still be shown so it
%% has to be manually commented out if you do not want it shown.
%%
%% Note that all of these author will be shown in the published article.
%% This feature is meant to be used prior to acceptance to make the
%% front end of a long author article more manageable.

\begin{abstract}
We discover the {significant} ({significance level of $\textgreater$99\%}) correlations between the fractional variation of the ionizing continuum and that of the C\,{\footnotesize IV} and/or Si\,{\footnotesize IV} BALs in each of 21 BAL quasars that have at least five-epoch observations from the Sloan Digital Sky Survey-I/II/III. This result reveals that the fluctuation of the ionizing continuum is the driver of {most} of these BAL variations. Among them, 17 show negative correlations and the other 4 positive correlations, which agrees with the prediction of photoionization models that absorption line variability response to ionization changes is not monotonic. 8 quasars out of 21 examples have been observed at least 30 times on rest-frame timescales as short as a few days, which reveals that changes in the incident ionizing continuum can cause BAL variability even in such a short period of time. In addition, we find that most of the 21 quasars show larger variation amplitude in Si\,{\footnotesize IV} than C\,{\footnotesize IV}, which reveals the ubiquity of saturation in these BALs (at least for C\,{\footnotesize IV} BALs).

\end{abstract}


%% subsection for cross--referencing in a \ref command.
%% cross--references to sections, equations, tables, and figures.


\section{Introduction} \label{sec:intro}

It is well known that the intrinsic absorption lines, including broad absorption lines (BALs; with line widths $\textgreater$2000 $\rm km\,s^{-1}$; \citealp{Weymann1991}), narrow absorption lines (NALs; with line widths of only a few hundred $\rm km\,s^{-1}$) and mini-BALs (with line widths between NALs and BALs; e.g., \citealp{Hamann2004}), can show variation in strength and shape over rest-frame timescales from days to years (e.g.,\citealp{Chen2015b,Grier2015,Wang2015,He2017,Chen2018a,Chen2018b,Hemler2019}, and references therein). The intrinsic absorption line variability may be caused by ionization state variability of the absorption gas (e.g., \citealp{Crenshaw2003}, and references therein) or by variability in the coverage fraction that caused by, for example, gas transverse motion (e.g., \citealp{Hamann2008}). Investigation of the relationship between the variability of absorption lines and that of the continuum can provide clues in understanding variation mechanism of the absorption lines.

In recent years, many efforts in systematic studies on the intrinsic absorption lines have been made to confirm the relationship between the variability of absorption lines and that of the continuum. In the field of BALs and/or mini-BALs, a few works have reported the lack of correlation between the BAL and the continuum  variations (e.g., \citealp{Gibson2008,Wildy2014,Vivek2014}). Based on a sample of 452 quasars from the Sloan Digital Sky Survey Data Release 10 (SDSS DR10), \citet{Wang2015} qualitatively confirmed that the variations of C\,{\footnotesize IV}, N\,{\footnotesize V}, and Si\,{\footnotesize IV} BALs/mini-BALs show high synchronization with those of the ionizing continuum and emission lines, and concluded that the variability of the absorption lines is driven mainly by variations in the gas ionization as response to the continuum variations. Based on the statistical analysis of a sample of 2005 BAL quasars from the SDSS-I/II/III, \citet{He2017} further concluded that variability of BAL in more than 80\% of quasars are caused by the variation of the ionizing continuum. More recently, correlation between the variation of the ionizing continuum and BALs has been confirmed by several papers (\citealp{LLQ2018,Lu2018saturation,Vivek2019}). In the field of NALs, \citet{Lu2017} confirmed, for the first time, the significant correlation between the variability of NALs and the ionizing continuum, using a two-epoch spectral sample of 40 quasars containing 52 variable C\,{\footnotesize IV}$\lambda$$\lambda$1548, 1551 absorption doublets constructed by \citet{Chen2015b}. \citet{Chen2018a, Chen2018b} further confirmed this correlation based on a larger variable NAL sample with more epoch observations.
%---------------------------------------------------------------------

\begin{splitdeluxetable*}{lccclrrrrBlrrrrclrrrr} 
\tabletypesize{\scriptsize}
\tablewidth{0pt} 
\tablenum{1}
\tablecaption{The properties of 21 quasars and correlation results. \label{tab.1}}
\tablehead{
\colhead{Quasar (SDSS J)} & \colhead{$z_{\rm em}$}& \colhead{$N_{\rm epochs}$} & \colhead{$N_{\rm pairs}$} & 
\multicolumn{5}{c}{$\Delta F_{\rm cont}/\langle F_{\rm cont}\rangle _{1350}$ vs. $\Delta \rm EW/\langle \rm EW \rangle_{C\,{\footnotesize IV}}$}&
\multicolumn{5}{c}{$\Delta F_{\rm cont}/\langle F_{\rm cont}\rangle _{1350}$ vs. $\Delta \rm EW/\langle \rm EW \rangle_{Si\,{\footnotesize IV}}$}&&
\multicolumn{5}{c}{$\Delta \rm EW/\langle \rm EW \rangle_{C\,{\footnotesize IV}}$ vs.  $\Delta \rm EW/\langle \rm EW \rangle_{Si\,{\footnotesize IV}}$} \\
\cline{5-14}\cline{16-20}
\colhead{} & \colhead{} &
\colhead{} & \colhead{} & \colhead{$r(p$-$value)^b$} & \colhead{$r_{\rm Bayes}^c$} & \colhead{$k^{\rm d}_{\rm Bayes}$} & \colhead{$b^{\rm e}_{\rm Bayes}$} &\colhead{$\sigma^{\rm f}_{\rm int}$} &
 \colhead{$r(p$-$value)^b$} & \colhead{$r_{\rm Bayes}^c$} & \colhead{$k^{\rm d}_{\rm Bayes}$} & \colhead{$b^{\rm e}_{\rm Bayes}$} &\colhead{$\sigma^{\rm f}_{\rm int}$} &&
 \colhead{$r(p$-$value)^b$} & \colhead{$r_{\rm Bayes}^c$} & \colhead{$k^{\rm d}_{\rm Bayes}$} & \colhead{$b^{\rm e}_{\rm Bayes}$} &\colhead{$\sigma^{\rm f}_{\rm int}$} 
} 
\colnumbers
\startdata 
022844.09+000217.0	&	2.720 	&	23	&	253	&	$	-0.37(0)	$	&	$	-0.37\pm	0.06	$	&	$	-0.26\pm	0.04 	$	&	$	0.07\pm	0.01 	$	&	0.01 	&	$	-0.42(0)	$	&	$	-0.40\pm	0.05 	$	&	$	-0.30\pm	0.04 	$	&	$	0.08\pm	0.01 	$	&	0.01 	&	&	$	0.83(0)	$	&	$	0.84\pm	0.02 	$	&	$	0.79\pm	0.03	$	&	$	0.00\pm	0.01 	$	&	0.00 	\\
023139.53+001758.3	&	2.382 	&	5	&	10	&	$	-0.78(0.0075)	$	&	$	-0.85\pm	0.18	$	&	$	-2.22\pm	0.74 	$	&	$	0.13\pm	0.08 	$	&	0.03 	&	$	-0.95(0)	$	&	$	-0.95\pm	0.11 	$	&	$	-9.90\pm	2.45 	$	&	$	0.14\pm	0.28 	$	&	0.20 	&	&	$	0.92(0.0002)	$	&	$	0.97\pm	0.07 	$	&	$	0.24\pm	0.04	$	&	$	0.10\pm	0.04 	$	&	0.01 	\\
023252.80-001351.1	&	2.030 	&	20	&	190	&	$	-0.34(0)	$	&	$	-0.32\pm	0.07	$	&	$	-0.40\pm	0.09 	$	&	$	-0.02\pm	0.02 	$	&	0.05 	&	$	-0.39(0)	$	&	$	-0.36\pm	0.07 	$	&	$	-1.26\pm	0.25 	$	&	$	-0.13\pm	0.06 	$	&	0.44 	&	&	$	0.92(0)	$	&	$	0.93\pm	0.01 	$	&	$	0.32\pm	0.01	$	&	$	0.03\pm	0.01 	$	&	0.01 	\\
024304.68+000005.4	&	2.002 	&	6	&	15	&	$	-0.59(0.021)	$	&	$	-0.68\pm	0.17	$	&	$	-0.26\pm	0.09 	$	&	$	-0.11\pm	0.03 	$	&	0.01 	&	$	-0.80(0.00034)	$	&	$	-0.77\pm	0.14 	$	&	$	-1.08\pm	0.28 	$	&	$	-0.05\pm	0.10 	$	&	0.10 	&	&	$	0.71(0.0028)	$	&	$	0.81\pm	0.12 	$	&	$	0.23\pm	0.05	$	&	$	-0.10\pm	0.03 	$	&	0.01 	\\
075007.63+275707.9	&	2.365 	&	5	&	10	&	$	-0.85(0.0016)	$	&	$	-0.98\pm	0.05	$	&	$	-0.83\pm	0.10 	$	&	$	0.03\pm	0.03 	$	&	0.00 	&	$	-0.87(0.0012)	$	&	$	-0.92\pm	0.10 	$	&	$	-2.01\pm	0.39 	$	&	$	0.10\pm	0.12 	$	&	0.03 	&	&	$	0.84(0)	$	&	$	0.96\pm	0.05 	$	&	$	0.39\pm	0.05	$	&	$	0.00\pm	0.03 	$	&	0.00 	\\
140554.87+530323.7$^a$	&	2.713 	&	32	&	496	&	$	0.03(0.57)	$	&	$	-0.02\pm	0.05	$	&	$	-0.02\pm	0.05 	$	&	$	-0.01\pm	0.01 	$	&	0.03 	&	$	-0.19(0)	$	&	$	-0.25\pm	0.05 	$	&	$	-0.33\pm	0.06 	$	&	$	0.07\pm	0.01 	$	&	0.05 	&	&	$	0.72(0)	$	&	$	0.71\pm	0.03 	$	&	$	0.54\pm	0.03	$	&	$	-0.05\pm	0.01 	$	&	0.01 	\\
141007.72+541203.6$^a$	&	2.335 	&	31	&	465	&	$	-0.56(0)	$	&	$	-0.55\pm	0.03	$	&	$	-0.93\pm	0.07 	$	&	$	-0.11\pm	0.01 	$	&	0.02 	&	$	-0.43(0)	$	&	$	-0.44\pm	0.04 	$	&	$	-0.72\pm	0.07 	$	&	$	-0.08\pm	0.01 	$	&	0.02 	&	&	$	0.83(0)	$	&	$	0.85\pm	0.01 	$	&	$	0.89\pm	0.03	$	&	$	-0.04\pm	0.00 	$	&	0.01 	\\
141421.54+522940.2$^a$	&	2.050 	&	32	&	496	&	$	-0.26(0)	$	&	$	-0.38\pm	0.04	$	&	$	-0.13\pm	0.01 	$	&	$	0.02\pm	0.00 	$	&	0.00 	&	$	-0.62(0)	$	&	$	-0.68\pm	0.03 	$	&	$	-0.29\pm	0.02 	$	&	$	0.00\pm	0.00 	$	&	0.00 	&	&	$	0.55(0)	$	&	$	0.73\pm	0.02 	$	&	$	0.57\pm	0.03	$	&	$	0.03\pm	0.00 	$	&	0.00 	\\
141955.28+522741.4$^a$	&	2.145 	&	32	&	496	&	$	-0.20(0)	$	&	$	-0.29\pm	0.06	$	&	$	-0.92\pm	0.22 	$	&	$	-0.02\pm	0.02 	$	&	0.05 	&	$	-0.40(0)	$	&	$	-0.60\pm	0.06 	$	&	$	-3.75\pm	0.50 	$	&	$	0.05\pm	0.03 	$	&	0.15 	&	&	$	0.87(0)	$	&	$	0.89\pm	0.01 	$	&	$	0.45\pm	0.01	$	&	$	0.00\pm	0.01 	$	&	0.01 	\\
142225.03+535901.9$^a$	&	2.691 	&	32	&	496	&	$	0.22(0)	$	&	$	0.35\pm	0.04	$	&	$	0.22\pm	0.03 	$	&	$	0.05\pm	0.00 	$	&	0.01 	&	$	0.06(0.18)	$	&	$	0.30\pm	0.04 	$	&	$	0.28\pm	0.04 	$	&	$	0.07\pm	0.01 	$	&	0.01 	&	&	$	0.56(0)	$	&	$	0.88\pm	0.01 	$	&	$	0.60\pm	0.02	$	&	$	0.01\pm	0.00 	$	&	0.00 	\\
142404.66+532949.6$^a$	&	2.768 	&	31	&	465	&	$	0.39(0)	$	&	$	0.47\pm	0.04	$	&	$	0.26\pm	0.03 	$	&	$	0.00\pm	0.00 	$	&	0.00 	&	$	-0.40(0)	$	&	$	0.47\pm	0.04 	$	&	$	0.45\pm	0.05 	$	&	$	0.02\pm	0.01 	$	&	0.01 	&	&	$	0.64(0)	$	&	$	0.70\pm	0.03 	$	&	$	0.40\pm	0.02	$	&	$	-0.01\pm	0.00 	$	&	0.00 	\\
142419.17+531750.8$^a$	&	2.533 	&	32	&	496	&	$	-0.36(0)	$	&	$	-0.52\pm	0.06	$	&	$	-0.38\pm	0.06 	$	&	$	-0.01\pm	0.00 	$	&	0.00 	&	$	-0.25(0)	$	&	$	-0.47\pm	0.07 	$	&	$	-2.28\pm	0.39 	$	&	$	-0.13\pm	0.03 	$	&	0.12 	&	&	$	0.67(0)	$	&	$	0.74\pm	0.03 	$	&	$	0.11\pm	0.01	$	&	$	0.00\pm	0.00 	$	&	0.00 	\\
142422.50+525903.3$^a$	&	2.142 	&	30	&	435	&	$	0.01(0.89)	$	&	$	0.08\pm	0.07	$	&	$	0.03\pm	0.02 	$	&	$	0.02\pm	0.00 	$	&	0.00 	&	$	-0.46(0)	$	&	$	-0.59\pm	0.05 	$	&	$	-1.06\pm	0.13 	$	&	$	-0.07\pm	0.02 	$	&	0.03 	&	&	$	0.33(0)	$	&	$	0.36\pm	0.04 	$	&	$	0.07\pm	0.01	$	&	$	0.03\pm	0.00 	$	&	0.00 	\\
164741.69+411545.3	&	2.082 	&	5	&	10	&	$	0.81(0.0049)	$	&	$	0.56\pm	0.27	$	&	$	0.91\pm	0.62 	$	&	$	0.26\pm	0.09 	$	&	0.05 	&	$	0.37(0.29)	$	&	$	0.38\pm	0.31 	$	&	$	0.54\pm	0.60 	$	&	$	0.26\pm	0.08 	$	&	0.05 	&	&	$	0.82(0.0038)	$	&	$	0.95\pm	0.07 	$	&	$	1.19\pm	0.17	$	&	$	-0.04\pm	0.05 	$	&	0.01 	\\
004323.43-001552.4	&	2.799 	&	7	&	21	&	$	-0.65(0.0016)	$	&	$	-0.71\pm	0.13	$	&	$	-3.04\pm	0.75 	$	&	$	-0.16\pm	0.19 	$	&	0.36 	&	$	-0.26(0.26)	$	&	$	-0.20\pm	0.22 	$	&	$	-0.83\pm	0.99 	$	&	$	-0.11\pm	0.25 	$	&	0.66 	&	&	$	0.85(0)	$	&	$	0.78\pm	0.10 	$	&	$	0.87\pm	0.17	$	&	$	-0.45\pm	0.13 	$	&	0.27 	\\
015048.83+004126.2	&	3.697 	&	5	&	10	&	$	-0.93(0.00011)	$	&	$	-0.88\pm	0.13	$	&	$	-1.18\pm	0.29 	$	&	$	0.03\pm	0.07 	$	&	0.01 	&	$	-0.94(0)	$	&	$	-0.98\pm	0.06 	$	&	$	-3.80\pm	0.58 	$	&	$	0.20\pm	0.15 	$	&	0.02 	&	&	$	0.93(0.00011)	$	&	$	0.92\pm	0.11 	$	&	$	0.33\pm	0.07	$	&	$	-0.03\pm	0.05 	$	&	0.01 	\\
022701.97-002621.7	&	2.295 	&	20	&	190	&	$	-0.60(0)	$	&	$	-0.57\pm	0.05	$	&	$	-1.20\pm	0.13 	$	&	$	0.00\pm	0.04 	$	&	0.29 	&	$	-0.74(0)	$	&	$	-0.73\pm	0.04 	$	&	$	-1.52\pm	0.12 	$	&	$	-0.06\pm	0.04 	$	&	0.20 	&	&	$	0.76(0)	$	&	$	0.78\pm	0.03 	$	&	$	0.79\pm	0.05	$	&	$	0.05\pm	0.03 	$	&	0.17 	\\
023233.57+000827.0	&	2.659 	&	19	&	171	&	$	-0.11(0.17)	$	&	$	-0.09\pm	0.09	$	&	$	-0.24\pm	0.24 	$	&	$	0.13\pm	0.04 	$	&	0.18 	&	$	-0.46(0)	$	&	$	-0.57\pm	0.07 	$	&	$	-2.18\pm	0.31 	$	&	$	0.29\pm	0.05 	$	&	0.26 	&	&	$	0.69(0)	$	&	$	0.77\pm	0.04 	$	&	$	0.53\pm	0.04	$	&	$	0.03\pm	0.02 	$	&	0.08 	\\
024557.23-000823.4	&	2.192 	&	5	&	10	&	$	0.89(0.00054)	$	&	$	0.80\pm	0.19	$	&	$	0.77\pm	0.27 	$	&	$	-0.05\pm	0.09 	$	&	0.03 	&	$	-0.25(0.49)	$	&	$	-0.27\pm	0.34 	$	&	$	-0.49\pm	0.83 	$	&	$	0.08\pm	0.28 	$	&	0.23 	&	&	$	-0.12(0.75)	$	&	$	-0.28\pm	0.44 	$	&	$	-0.28\pm	8.24	$	&	$	0.12\pm	0.24 	$	&	0.07 	\\
024747.59-004810.0	&	1.913 	&	5	&	10	&	$	-0.90(0.00034)	$	&	$	-0.07\pm	0.59	$	&	$	-3.12\pm	*^{\rm g}$$	$	&	$	-0.12\pm	*^{\rm g}	$	&	0.19 	&	$	-0.61(0.06)	$	&	$	0.04\pm	0.53 	$	&	$	*^{\rm g}		$	&	$	*^{\rm g}		$	&	0.14 	&	&	$	0.65(0.043)	$	&	$	0.49\pm	0.28 	$	&	$	0.71\pm	0.53	$	&	$	0.11\pm	0.22 	$	&	0.22 	\\
234315.88+004659.5	&	2.775 	&	5	&	10	&	$	-0.77(0.0092)	$	&	$	-0.74\pm	0.25	$	&	$	-0.13\pm	0.07 	$	&	$	0.00\pm	0.02 	$	&	0.00 	&	$	-0.76(0.011)	$	&	$	-0.84\pm	0.20 	$	&	$	-0.86\pm	0.34 	$	&	$	-0.11\pm	0.10 	$	&	0.01 	&	&	$	0.52(0.13)	$	&	$	0.44\pm	0.31 	$	&	$	0.09\pm	0.08	$	&	$	0.02\pm	0.02 	$	&	0.00 	\\
\enddata
\begin{tablenotes}
\footnotesize
\item$^{\rm a}$Quasars of the SDSS-RM project.
\item$^{\rm b}$The Spearman rank correlation coefficient. The values in brackets are $p$-values, whose zero values represent $p$-value$\textless$0.0001.
\item$^{\rm c}$The correlation coefficient from Bayesian approach \citep{Kelly2007}.
\item$^{\rm d}$The slope of the linear fit.
\item$^{\rm e}$The constant in the regression.
\item$^{\rm f}$The variance of the intrinsic scatter.
\item$^{\rm g}$The outliers that greater than $10^3$, which mainly due to the large errors of the fraction variation of the continuum of SDSS J024747.59--004810.0.

\end{tablenotes}
\end{splitdeluxetable*}
%---------------------------------------------------------------------

Although the correlation between the the variability of intrinsic absorption lines and that of the ionizing continuum has been confirmed based on statistical analyses, the relationships between them in individual quasars are still unclear. As far as we know, only a few studies have reported correlation analysis between the variability of intrinsic absorption lines and that of the ionizing continuum in individual quasars with multi-epoch observations (\citealp{Gabel2005,He2014,Huang2019}). The large number of quasars with multi-epoch spectroscopic from the SDSS (\citealp{York2000}) provide a good opportunity for the research in this field. In this paper, we will investigate the relationships between the variability of absorption lines and that of the continuum in individual quasars with multi-epoch spectroscopic observations by SDSS. The sample selection and correlation analysis methods are given in Section \ref{sec:2}. The discussions about our results are presented in Section \ref{sec:disscu}. A brief conclusion is given in Section \ref{sec:Conclusion}.

%----------------------------------------------------

\section{Sample selection and correlation analysis} \label{sec:2} %2%%%%%%%%%%%
\begin{figure*}\centering
\includegraphics[width=1.5\columnwidth]{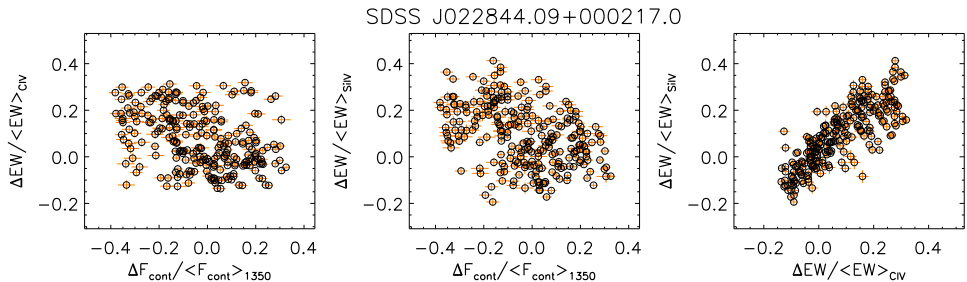}
%\plotone{f1.eps}[\textwidth=\columnwidth]
\caption{Plots showing the $\Delta \rm EW/\langle \rm EW \rangle_{C\,{\footnotesize IV}}$ vs. $\Delta F_{\rm cont}/\langle F_{\rm cont}\rangle _{1350}$ (left panels), $\Delta \rm EW/\langle \rm EW \rangle_{Si\,{\footnotesize IV}}$ vs. $\Delta F_{\rm cont}/\langle F_{\rm cont}\rangle _{1350}$ (middle panels), and $\Delta \rm EW/\langle \rm EW \rangle_{Si\,{\footnotesize IV}}$ vs. $\Delta \rm EW/\langle \rm EW \rangle_{C\,{\footnotesize IV}}$ (right panels) for each of 21 quasars. \label{fig.1}}
\end{figure*}

Our initial sample consists of 2005 BAL quasars constructed by \citet{He2017} from the SDSS-I/II/III. Among them, there are 46 quasars have at least five-epoch observations and a signal-to-noise ratio (S/N) level of S/N$\textgreater$10 in one-epoch observation at least. Because each quasar has $m$ ($m\geq5$) spectra, there are a total of $C^2_m=m(m-1)/2$ spectra pairs. We downloaded spectra of these 46 quasars from SDSS DR14 and fitted power-law continua for them using the procedure from \citet{LLQ2018}. We used the power-law continuum flux at 1350\,\AA~to represent the strength of the ionizing continuum for each spectrum. The equivalent width (EW) values of the BALs were derived from the catalog of \citet{He2017}. Then we performed the correlation analyses between the fractional variation of the ionizing continuum and that of the C\,{\footnotesize IV} and/or Si\,{\footnotesize IV} BALs for each two observations of each the 46 quasars. Equations for calculating the fractional variation of the ionizing continuum and BALs is adopted from \citet{LLQ2018} (equations (2) and (4) in therein). Finally, we got 21 out of 46 quasars showing significant correlations (p-value$\textless$0.01) between the fractional variation of the ionizing continuum and that of the C\,{\footnotesize IV} and/or Si\,{\footnotesize IV} BALs. Plots of the fractional variation of the ionizing continuum with that of the C\,{\footnotesize IV} and Si\,{\footnotesize IV} BALs, as well as the C\,{\footnotesize IV} versus Si\,{\footnotesize IV} BALs are shown in Figure \ref{fig.1}. The results of the Spearman’s rank correlation analysis are listed in Table \ref{tab.1}. 

%\Delta F_{\rm cont}/\langle F_{\rm cont}\rangle _{1350}$}  & & \multicolumn2c{$\Delta F_{\rm cont}/\langle F_{\rm cont}\rangle _{1350}$}& & \multicolumn2c{$\Delta \rm EW/\langle \rm EW \rangle_{C\,{\footnotesize IV}}$}  & & &&\multicolumn2c{vs.}& &\multicolumn2c{vs.}&&\multicolumn2c{vs.}    & & &&\multicolumn2c{$\Delta \rm EW/\langle \rm EW \rangle_{C\,{\footnotesize IV}}$}& &\multicolumn2c{$\Delta \rm EW/\langle \rm EW \rangle_{Si\,{\footnotesize IV}}$} &&\multicolumn2c{$\Delta \rm EW/\langle \rm EW \rangle_{Si\,{\footnotesize IV}}$} \\

\begin{figure*}\centering
\includegraphics[width=1.5\columnwidth]{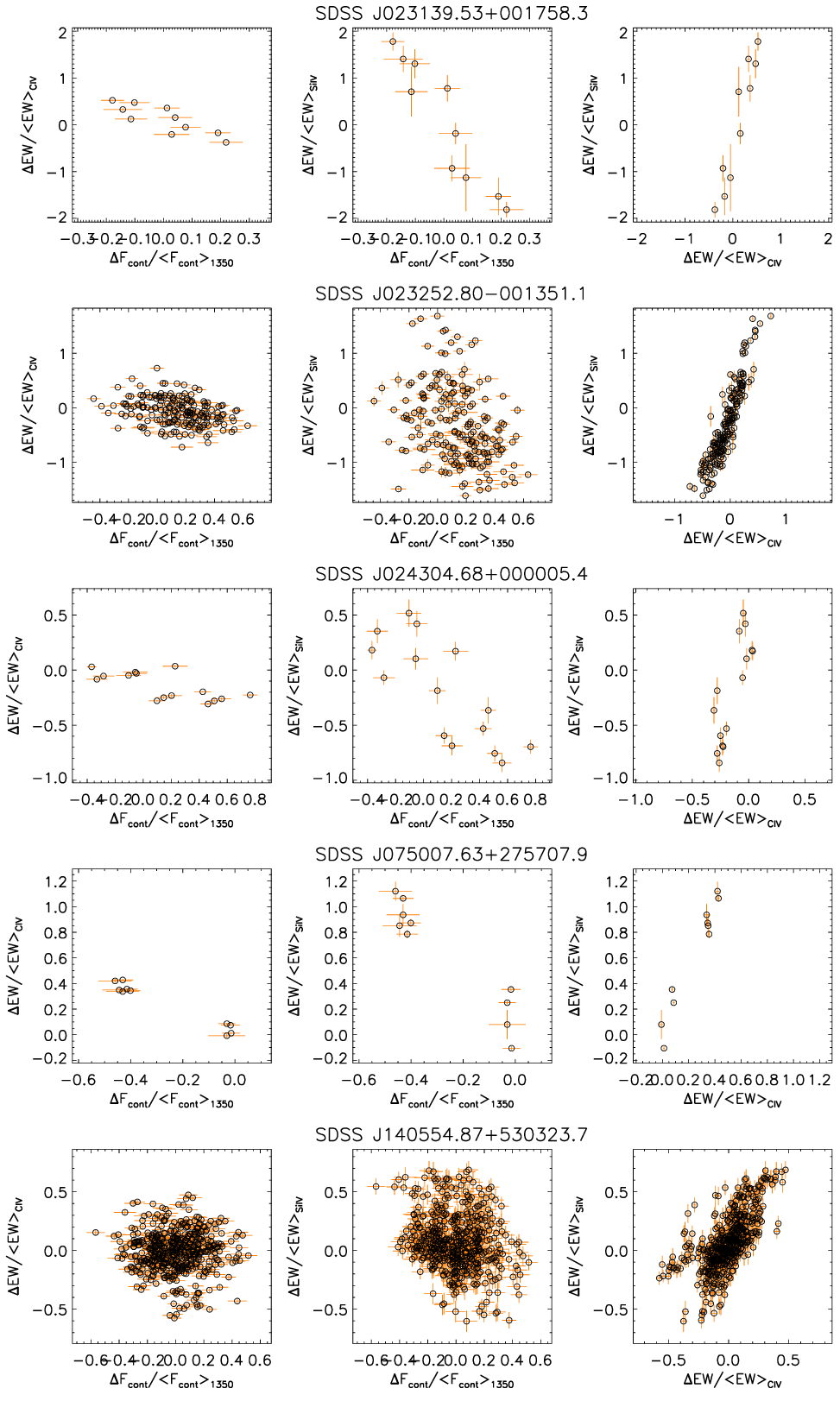}

%\plotone{f1.eps}[\textwidth=\columnwidth]
\caption*{Figure 1 $-continued$}  
\end{figure*}
\begin{figure*}\centering

\includegraphics[width=1.5\columnwidth]{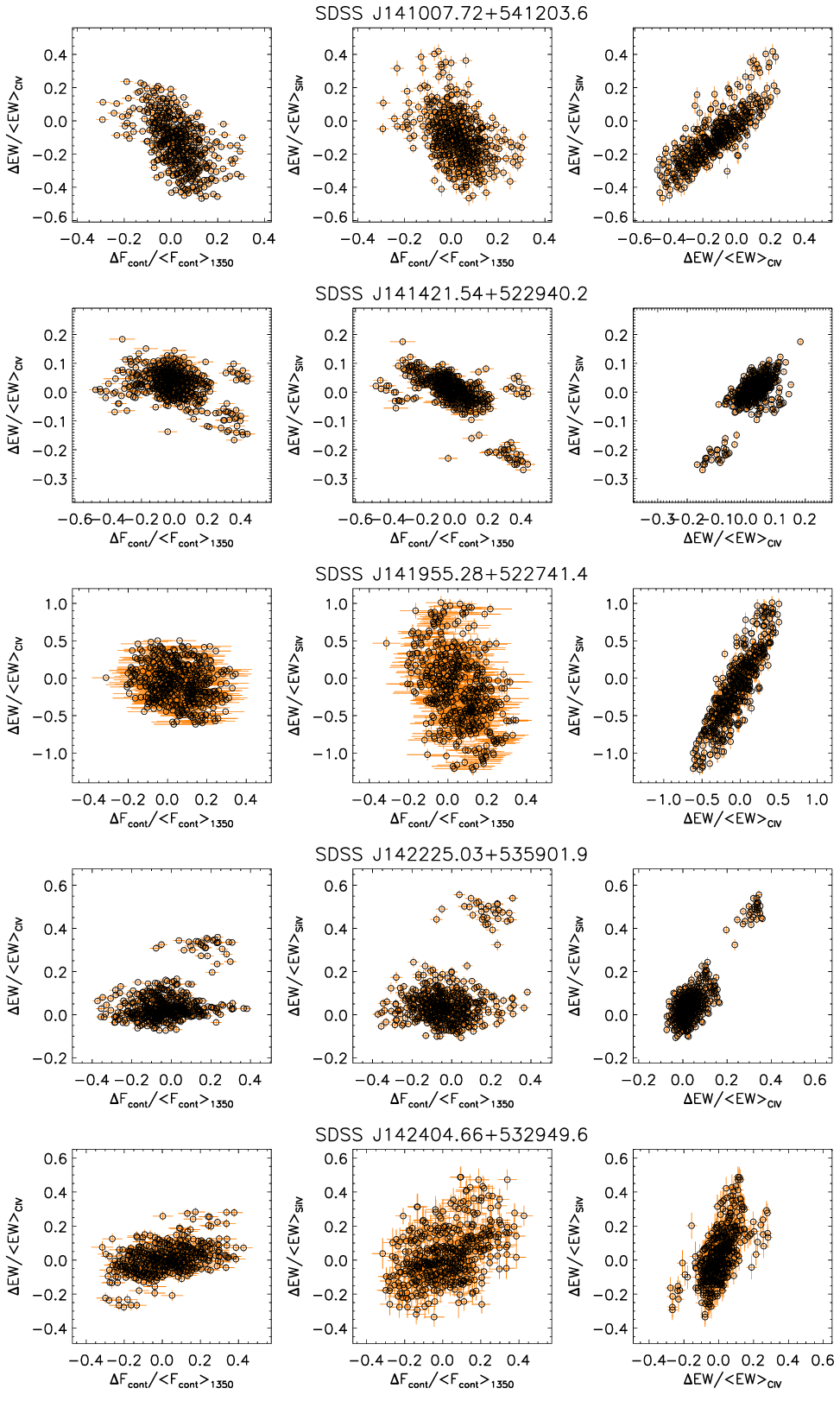}

%\plotone{f1.eps}[\textwidth=\columnwidth]
\caption*{Figure 1 $-continued$}  
\end{figure*}
\begin{figure*}\centering

\includegraphics[width=1.5\columnwidth]{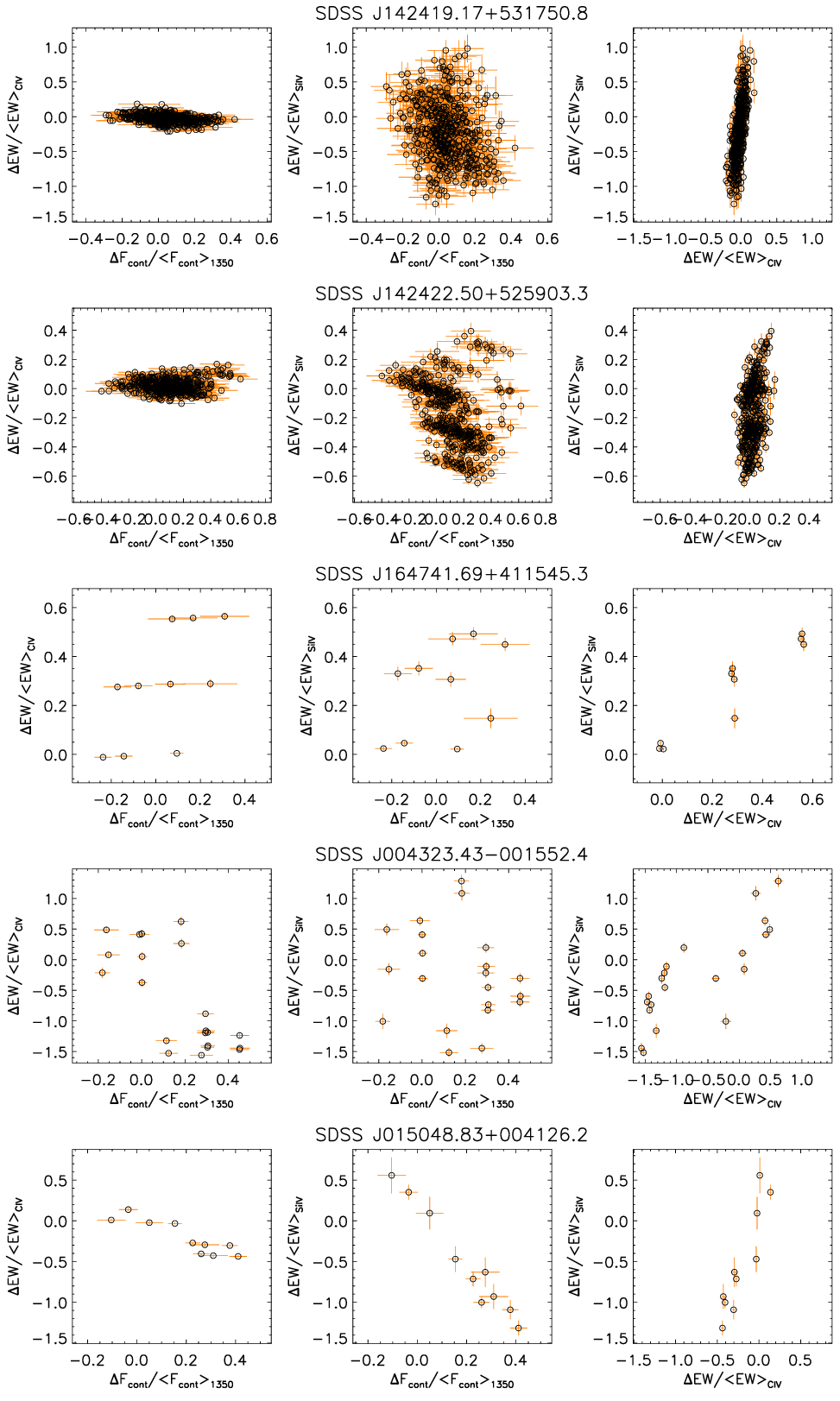}

%\plotone{f1.eps}[\textwidth=\columnwidth]
\caption*{Figure 1 $-continued$}
\end{figure*}

\begin{figure*}\centering

\includegraphics[width=1.5\columnwidth]{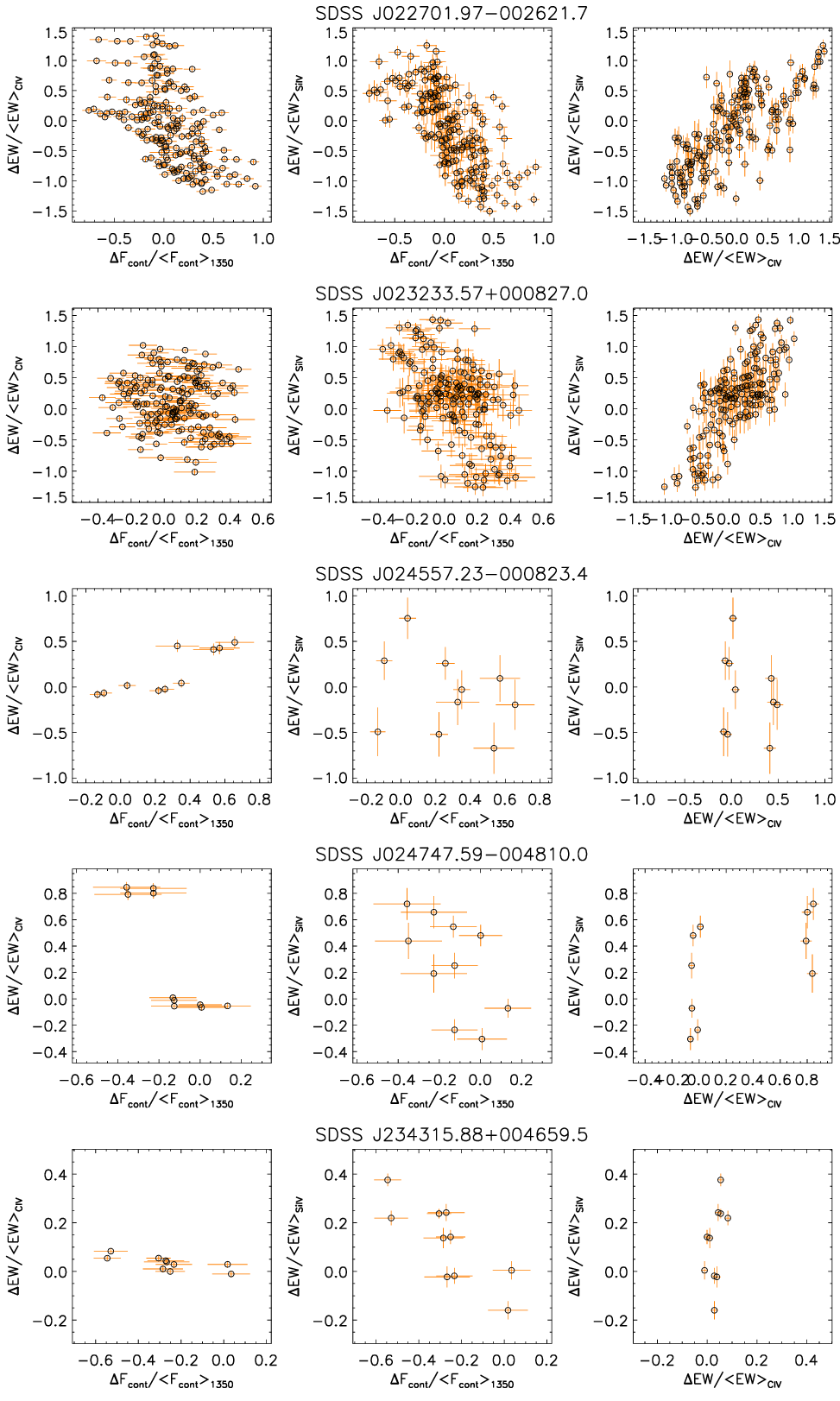}

%\plotone{f1.eps}[\textwidth=\columnwidth]
\caption*{Figure 1 $-continued$} 
\end{figure*}

\section{Results and Discussions} \label{sec:disscu}%3%%%%%%%%%%
\subsection{The cause of the BAL variability} \label{sec:cause} %3.1%%%%%%%%%%%
The most important result in this paper is that we confirm the significant correlations between the fractional variation of the ionizing continuum and that of the C\,{\footnotesize IV} and/or Si\,{\footnotesize IV} BALs for each of 21 quasars. This is the first systematic study of the correlation analysis between the variability of intrinsic absorption lines and that of the ionizing continuum in individual quasars based on a large sample with multi-epoch observations. These statistical results serve as strong evidence for the idea that fluctuation of the ionizing continuum is the driver of {most} of the BAL variation of these quasars. 

Among these sources, two have been investigated in details by previous papers. For one of them, SDSS J022844.09+000217.0 has been investigated by \citet{He2014}. They found no significant correlation between the EW of C\,{\footnotesize IV}/Si\,{\footnotesize IV} BAL trough and the ionizing continuum in this source based on 18 epochs of spectroscopic observations, so they concluded that the BAL-trough variability is not dominated by photoionization. However, we do find significant moderate anticorrelation between the fractional variation of the ionizing continuum and that of both C\,{\footnotesize IV} and Si\,{\footnotesize IV} BALs (see Table \ref{tab.1}), when combining with five more epochs of spectroscopic observations (Table \ref{tab.1}). Based on our results, we hold the view that the BAL-trough variation in SDSS J022844.09+000217.0 is dominated by photoionization, which is the response to the change of the ionizing continuum. 

For the other one, SDSS J141007.74+541203.3 has been severally investigated by \citet{Grier2015} and \citet{Huang2019}. \citet{Grier2015} found that (1) variations over the whole BAL trough rather than in some snippets, and (2) coordinate variability of the high-velocity C\,{\footnotesize IV} BAL (their Trough A) and the mini-BAL (their Trough B). \citet{Huang2019} reported the anticorrelation between the EW of BALs and the flux of the ionizing continuum of this quasar, based on just the spectra showing significant EW variations. Both \citet{Grier2015} and \citet{Huang2019} held the view that the most likely cause for the BAL variability in this source is a rapid response to the ionizing continuum changes. In this paper, we confirmed the significant anticorrelation between the fractional variation of the ionizing continuum and that of both C\,{\footnotesize IV} and Si\,{\footnotesize IV} BALs (see Table \ref{tab.1}), which is consistent with the conclusion made by \citet{Grier2015} and \citet{Huang2019}.

\subsection{Positive vs. negative correlations} \label{sec:Positive vs. negative} %3.2%%%%%%%%%%%
Another important result of our study is that we found both negative and positive correlation cases between the fractional variation of the ionizing continuum and that of the C\,{\footnotesize IV} and/or Si\,{\footnotesize IV} BALs. In fact, both the photoionization simulations (e.g., \citealp{Wang2015,He2017}) and systematic studies of the absorption lines (\citealp{Wang2015,He2017,Lu2017,Chen2018b}) have heralded that the ionic column density of a specific species could show response to a continuum variability positively or negatively. On the one hand, photoionization simulations show that absorption line response to ionization variability is not monotonic (see e.g., figure 3 in \citealp{He2017}). When the absorbing gas are at a relatively low ionization state, the EW of the C\,{\footnotesize IV}/Si\,{\footnotesize IV} BAL show positive response to the continuum variation, otherwise, the negative response reveals the relatively high ionization state. On the other hand, based on large absorption line samples, statistical studies in both BALs (e.g., \citealp{Wang2015,He2017}) and NALs (\citealp{Lu2017,Chen2018b}) also showed that the variations of intrinsic absorption lines are positively or negatively responded to the alternations of quasar continua. These two kinds of respond suggest that the variable intrinsic absorption lines can be divided into at least two classes: one is highly ionized systems dominated by absorbing gas in high ionization and show negative response to changes of quasar emissions; and the other is lowly ionized systems dominated by absorbing gas in low ionization and show positive response to changes of quasar emissions. 

However, rare convincing individual cases on the EW of C\,{\footnotesize IV}/Si\,{\footnotesize IV} BAL showing positive response to the continuum variation have been reported. Here, we find that 17 out of 21 examples show negative correlations, while 4 (SDSS J142225.03+535901.9, SDSS J142404.66+532949.6, SDSS J164741.69+411545.3, SDSS J024557.23--000823.4) out of 21 examples show positive correlations. Our results are consistent with the previous systematic studies and provide observational evidence for the photoionization model in explaining the variation of BALs.

\subsection{Rapid BAL variability} \label{sec:Rapid} %3.3%%%%%%%%%%%
8 out of our 21 examples were intensively observed in the year of 2014 in the Sloan Digital Sky Survey Reverberation Mapping Project (SDSS-RM; \citealp{Shen2015a}), which obtained spectra of 32 epochs of 849 quasars on the SDSS 2.5m telescope (\citealp{Gunn2006,Smee2013}). In this project, the median spacing between observations is as short as about 4 days (\citealp{Shen2015a}). Analyses on the time variability of BALs of the SDSS-RM project were reported by \citet{Grier2015} and \citet{Hemler2019}. In this paper, we confirmed the significant correlation between the fractional variation of the ionizing continuum and that of C\,{\footnotesize IV} and/or Si\,{\footnotesize IV} BALs (see Table \ref{tab.1}) in 8 quasars of the SDSS-RM project, which serve as strong evidence for ionization driven BAL variability as rapid response to changes in the incident ionizing continuum.

\subsection{{Saturation effect}} \label{sec:3.4} %3.4%%%%%%%%%%%
Saturation in absorption lines can make significant influence on the correlation between the continuum and absorption line variations because a saturated absorber might respond to the continuum fluctuations {softly} or even no respond. \citet{LLQ2018} pointed out that saturation in absorption lines could be one of the reasons for the substantial scatter of the plots for the fractional variation of the ionizing continuum versus that of both C\,{\footnotesize IV} and Si\,{\footnotesize IV} BALs. \citet{LLQ2018} also found that the fractional variations of Si\,{\footnotesize IV} BALs seem greater than C\,{\footnotesize IV} BALs, which indicates that the C\,{\footnotesize IV} BALs suffering from more saturation than Si\,{\footnotesize IV} BALs. \citet{Lu2018saturation} further confirmed the moderate anticorrelation between the Si\,{\footnotesize IV} BAL fractional variations and the ionizing continuum in 74 quasars that show obvious change in Si\,{\footnotesize IV} BAL but no or small change in C\,{\footnotesize IV} BAL (hereafter Phenomenon I), revealing the ubiquitous effect of the ionizing continuum variability on Phenomenon I. \citet{Vivek2019} found that the BAL sample with shallow trough shows stronger correlation between the absorption line and the ionizing continuum variability. 

Visually check Figure \ref{fig.1} we can find that most of the quasars show larger variation amplitude in Si\,{\footnotesize IV} than C\,{\footnotesize IV}, which is in agreement with previous studies. In particular, several quasars (e.g., SDSS J023139.53+001758.3, SDSS J142419.17+531750.8 and SDSS J142422.50+525903.3) show Phenomenon I. These results reveal the ubiquity of saturation in these BALs (at least for C\,{\footnotesize IV} BALs), however, the ionization variation in response to the ionizing continuum variations could still be the driver of their changes. 
\subsection{{Limitations}} \label{sec:limitation} %3.5%%%%%%%%%%%
{Although the significant correlations between the fractional variation of the ionizing continuum and that of the C\,{\footnotesize IV} and/or Si\,{\footnotesize IV} BALs in each of 21 BAL quasars have been confirmed by the Spearman’s rank correlation tests, most of the relations show substantial intrinsic scatter, which is corroborated by small correlation coefficients. Excepting the saturation effect from the BAL troughs (Section \ref{sec:3.4}), the quality of the SDSS spectra (both resolution and S/N) may have a significant effect on this scatter. As a consistency check, we estimated the correlation coefficients and {performed linear fittings} using the Bayesian approach by \citet{Kelly2007}, which takes into account the measurement errors in the variables and intrinsic scatter. As shown in Table \ref{tab.1}, most of the Bayesian results are consistent with the results of the Spearman’s rank correlation analysis, which indicates that most of the reported correlations indeed hold. The only exception is the SDSS J024747.59--004810.0, which shows big deviation with large error. Future high quality monitoring will help us to confirm whether the correlation indeed hold or not in this quasar.} 

{Another reason for the scatter could be that some quasars actually show no obvious EW variation in most of their observations. For example, the SDSS J142225.03+535901.9 and SDSS J141421.54+522940.2 both show obvious EW variation in only a few observations, which could be the reason of the gaps shown in the plots.}

\section{{Conclusion}} \label{sec:Conclusion} %4%%%%%%%%%%%
We have made the correlation analyses between the fractional variation of the ionizing continuum and that of the C\,{\footnotesize IV} and Si\,{\footnotesize IV} BALs for each of 46 BAL quasars that have been observed by SDSS at least 5 times, and have confirmed {significant} correlations in 21 of them. We have presented the following results and discussions on them.

(1) The significant correlations between the fractional variation of the ionizing continuum and that of the C\,{\footnotesize IV} and/or Si\,{\footnotesize IV} BALs for each of 21 quasars have been found. We think this result reveals the fluctuation of the ionizing continuum is the driver of {most} of these BAL variations.

(2) We find that 17 out of them show negative correlations, while the other 4 examples show positive correlations. These results support the previous systematic studies and the photoionization model, in which when the absorbing gas are at a relatively low ionization state, the EW of the C\,{\footnotesize IV}/Si\,{\footnotesize IV} BAL shows positive response to the continuum variation, otherwise, a negative response appears. 

(3) 8 quasars out of our sample have been observed at least 30 times within a few days. The significant correlations between the fractional variation of the ionizing continuum and that of C\,{\footnotesize IV} and/or Si\,{\footnotesize IV} BALs (see Table \ref{tab.1}) in these quasars serve as strong evidence for ionization driven BAL variability as rapid response to changes in the incident ionizing continuum.

(4) We find that most of the 21 quasars show larger variation amplitudes in Si\,{\footnotesize IV} than C\,{\footnotesize IV}, which reveals the ubiquity of saturation in these BALs (at least for C\,{\footnotesize IV} BALs). Saturation in BALs may lead to the substantial dispersion of the plots for the fractional variation of BALs. Even so, the correlations between the absorption line and the ionizing continuum variability still be apparently presented.

%----------------\section*{Acknowledgements}--------------------------
\acknowledgments
We are very grateful to the anonymous referee and the statistics editor for comments that improved the quality of this article. 

Funding for the Sloan Digital Sky Survey IV was
provided by the Alfred P. Sloan Foundation, the U.S.
Department of Energy Office of Science, and the Participating
Institutions. SDSS-IV acknowledges support and resources
from the Center for High-Performance Computing at the
University of Utah. The SDSS website is \url{http://www.sdss.org/}.

SDSS-IV is managed by the Astrophysical Research
Consortium for the Participating Institutions of the SDSS
Collaboration including the Brazilian Participation Group, the
Carnegie Institution for Science, Carnegie Mellon University,
the Chilean Participation Group, the French Participation
Group, Harvard-Smithsonian Center for Astrophysics, Instituto
de Astrofísica de Canarias, The Johns Hopkins University,
Kavli Institute for the Physics and Mathematics of the Universe
(IPMU)/University of Tokyo, Lawrence Berkeley National
Laboratory, Leibniz Institut für Astrophysik Potsdam (AIP),
Max-Planck-Institut für Astronomie (MPIA Heidelberg),
Max-Planck-Institut für Astrophysik (MPA Garching), MaxPlanck-Institut für Extraterrestrische Physik (MPE), National
Astronomical Observatories of China, New Mexico State
University, New York University, University of Notre Dame,
Observatário Nacional/MCTI, The Ohio State University,
Pennsylvania State University, Shanghai Astronomical Observatory, United Kingdom Participation Group, Universidad
Nacional Autónoma de México, University of Arizona,
University of Colorado Boulder, University of Oxford,
University of Portsmouth, University of Utah, University of
Virginia, University of Washington, University of Wisconsin,
Vanderbilt University, and Yale University.

\bibliographystyle{aasjournal}%plain \bibliographystyle{mnras}%try
\bibliography{NALvsBAL} % if your bibtex file is called example.bib

\end{document}